\documentclass[12pt]{article}
\usepackage{epsf,epsfig,psfig,float,amssymb,latexsym}
\newcommand{\NP}[1]{ Nucl.\ Phys.\ {\bf #1}}
\newcommand{\ZP}[1]{ Z.\ Phys.\ {\bf #1}}

\newcommand{\PL}[1]{ Phys.\ Lett.\ {\bf #1}}

\newcommand{\AN}[1]{Ann. Phys. NY {\bf #1}}

\newcommand{\PR}[1]{Phys.\ Rev.\ {\bf #1}}
\newcommand{\PRL}[1]{ Phys.\ Rev.\ Lett.\ {\bf #1}}

\newcommand{\IJMP}[1]{ Int. J. Mod. Phys.\ {\bf #1}}
\newcommand{\EPJ}[1]{Eur. Phys. J.\ {\bf #1}}
\newcommand{\bi}{\bibitem}
\newcommand{\vs}{\vspace{-0.20cm}}
\newcommand{\Opu}{\mathcal{O}(p)}
\newcommand{\Opd}{\mathcal{O}(p^2)}
\newcommand{\Opt}{\mathcal{O}(p^3)}
\newcommand{\Opc}{\mathcal{O}(p^4)}

\newcommand{\be}{\begin{equation}}
\newcommand{\ee}{\end{equation}}
\newcommand{\ba}{\begin{eqnarray}}
\newcommand{\ea}{\end{eqnarray}}

\newcommand{\barr}[1]{\not\mathrel #1}

\begin{document}

\hfill \small{FZJ-IKP(TH)-1999-36}

\thispagestyle{empty}

\vspace{2cm}

\begin{center}
{\Large{\bf Chiral unitary meson--baryon dynamics \\[0.3em] 
   in the presence of resonances:\\[0.6em]
    Elastic pion--nucleon scattering
}}
\end{center}
\vspace{.5cm}

\begin{center}
{\large{Ulf-G. Mei{\ss}ner\footnote{email: 
                           Ulf-G.Meissner@fz-juelich.de} and 
  J. A. Oller\footnote{email: j.a.oller@fz-juelich.de}}}
\end{center}

\begin{center}
{\it {\it Forschungzentrum J\"ulich, Institut f\"ur Kernphysik (Theorie) \\ 
D-52425 J\"ulich, Germany}}
\end{center}
\vspace{1cm}

\begin{abstract}
\noindent
{\small{We develop a novel approach to chiral meson--baryon dynamics
incorporating unitarity constraints and explicit resonance fields.
It is based on the most general structure of any pion-nucleon partial
wave amplitude neglecting the unphysical cuts as derived from the
$N/D$ method. This amplitude is then matched
to the one--loop heavy baryon chiral perturbation theory result at
third order and to tree level exchanges of baryon-- and meson
states in the $s,t$ and $u$ channels. This generates the
left--hand cuts. The unitarization procedure
does not involve form factors or
regulator functions. The resonance parameters
are determined from fits to the S-- and P--wave pion--nucleon partial 
wave amplitudes for energies up to 1.3~GeV. In particular, the $\Delta
(1232)$ is accurately reproduced  whereas scalar and vector meson
couplings are less precisely pinned down. We also obtain a
satisfactory description of the $\pi$N threshold parameters. Further 
extensions of this method to coupled channels and the three--flavor
case are briefly discussed.
}}  
\end{abstract}

\vspace{2cm}

\begin{center}
\end{center}

\newpage

\section{Introduction}
\def\theequation{\arabic{section}.\arabic{equation}}
\setcounter{equation}{0}

There are many reasons why it is interesting to study pion-nucleon
scattering. First, in the threshold region, chiral perturbation theory
is applicable and thus the chiral structure of QCD can be
investigated, see e.g.~\cite{nadia} (and references therein). 
This very precise method is, however, only applicable
in the threshold and in the unphysical region since there the pion
momenta are small. At higher energies, additional physics due to the
appearance of resonances and coupled channel effects sets in. In fact,
the dispersive analysis of pion--nucleon scattering data has been and
still is the best method to analyze the properties of the baryon
resonances~\cite{Hohler} (as long as they couple strongly to the $\pi$N
system). Many models and approaches of different types have been
developed to deal with these phenomena, we just mention the
meson--exchange~\cite{juelichpin} and the extended tree level~\cite{tree}
models. While all these
models in general give a good description of the phase shifts (and
inelasticities), they are not including the pion loops resulting from
the chiral structure of QCD in a systematic fashion (the often
employed unitarization schemes or resummation techniques include some
classes of these diagrams, but not all). Furthermore, in these 
references most of the tree level diagrams involving the exchange of meson 
resonances do not have the  momentum dependence  required by 
chiral symmetry. Consequently, it would be
interesting to devise a scheme that at low energies exactly reproduces
the chiral perturbation theory amplitudes (to a given order in the
chiral expansion) but also allows one to work in the resonance region, including 
also resonance exchange at tree level in conformity with the 
non--linearly realized chiral symmetry. 
Such an approach will be developed here. It is similar to the one 
set up in ref.\cite{nd} for meson--meson scattering and will be
explained in more detail below. Before doing that, it is worth to
stress that this method can easily be extended to the coupled channel
case and also to the three--flavor sector, in particular also for the
description of kaon--nucleon scattering. While there have been
already some successful studies of chiral SU(3) dynamics with coupled channels
for this and related processes~\cite{norbert,valencia}, no attempt has
been made so far to match the $\pi$N subprocesses to the accurate
chiral perturbation theory predictions. Thus, the model presented here
can be considered as a first building block for setting up a relativistic 
coupled channel scheme which fulfills these requirements.

\medskip\noindent
It is known that dispersion theory can be used to match on chiral
perturbation theory and thus allows one to extend the range of 
energies one can consider. Of course, any unitarization scheme is at
variance with the power counting underlying an effective field theory
like e.g. CHPT. It is, however, not known how to extend such a concept
to include resonances (with the exception of the $\Delta (1232)$
isobar, see ref.\cite{HHK}). On the other hand, unitarization methods
allow one to exactly fulfill unitarity (which is only respected
perturbatively in CHPT) and can generate S--matrix poles corresponding
to resonances. We attempt to combine these concepts in a way that
makes best use of these various features. For that, employing the
well--known $N/D$ method~\cite{chew}, we derive the most general
structure for any $\pi$N partial wave amplitude when the unphysical
cuts are neglected. Then, similarly as in ref. \cite{WW}, the unphysical cuts 
will be treated as a perturbation in a loop expansion with the HBCHPT power 
counting. Hence, our starting point of no unphysical cuts at all is the 
zeroth order approach in this scheme from which the first order approach, 
the one necessary to match with $\Opt$ HBCHPT, is developed. Note that 
tree level diagrams involving local operators and particle exchanges 
in all the channels are also included since they do not involve
loops. This should be a good starting point for the channels which are
dominated by unitarity and the presence of $s$--, $u$-- or $t$--channel
resonances, like e.g. the $P_{33}$ partial wave which is largely given
by the $\Delta (1232)$. The argument is further strengthened by the 
observation made in ref.\cite{aspects}, namely that the dimension
two low--energy constants of the chiral effective $\pi$N Lagrangian
are saturated by $s$--, $u$-- and $t$--channel resonance contributions, most
notably the $\Delta$ and the $\rho (770)$. In addition, for the
scalar--isoscalar channel one can deduce a certain contribution from 
a meson with these quantum numbers. In ref.\cite{aspects}, this meson
was taken to be a light $\sigma$, but in fact only the ratio of the
scalar meson--nucleon coupling constant to the scalar meson mass
plays a role. We will come back to this point in much more detail
below. In addition, we will also include the Roper $N^* (1440)$
resonances which plays a minor but non--negligible role in the energy
range we will consider (since we do not calculate inelasticities, we
stay below center--of--mass energies of about 1.3~GeV). Note that 
preferably we should match to a relativistic $\pi$N chiral
perturbation theory amplitude because in that way one can avoid
the difficulties with the HBCHPT approach as detailed in
ref.\cite{BL}. Such an amplitude, based on a proper
power counting, is not yet available but under construction, see
ref.\cite{BL}. It remains to be seen whether the matching to such an
amplitude will lead to better results than the ones presented here.

\medskip\noindent
Our manuscript is organized as follows. In section~2, we collect the
formalism necessary to describe elastic pion--nucleon scattering.
Section~3 we construct the basic tree level diagrams supplemented by the 
$\Opt$ HBCHPT loops. In section~4 the unitarization method is discussed and
in sections 5 the results are presented. The most relevant
conclusions of the work are summarized in section 6.

\section{Prelude: Kinematics}
\def\theequation{\arabic{section}.\arabic{equation}}
\setcounter{equation}{0}
\label{sec:kine}
In this section, we collect the necessary formalism concerning elastic pion--nucleon
scattering. This serves to set our notation and to keep the manuscript 
self--contained. In the following, we consider the $\pi$N scattering amplitude. 
In the center--of--mass frame (c.m.), the amplitude for the process 
$\pi(q_1)\,+\,N(q_2) \rightarrow \pi(q_2)\, +\, N(p_2)$ can be written as
(suppressing isospin indices)
\begin{equation}
\label{spinflip}
T(W,t)=\left( \frac{E+m}{2m} \right) \biggl[ g(W,t)+i\vec{\sigma}\cdot
(\vec{q}_2  \times \vec{q}_1 ) h(W,t) \biggr]~,
\end{equation}
where $\vec{q}_1$ and $\vec{q}_2$ are the three-momenta of the incoming and
the outgoing pion, respectively, $W\equiv \sqrt{s}$ is the total
c.m. energy, $m$ is the nucleon mass and
$t=(q_1-q_2)^2$ is the invariant momentum transfer squared. 
Furthermore, $g(W,t)$ refers to the non-spin-flip amplitude and
$h(W,t)$ to the spin-flip one.
For a given total isospin ($I$) of the $\pi$N system, the partial 
wave amplitudes $T_{l\pm}^I$, where $l$
refers to the orbital angular momentum and the subscript $\pm$ to the total
angular momentum ($J=l\pm 1$), are given in terms of the functions 
$g(W,t)$ and $h(W,t)$ via
\begin{equation}
\label{pw}
T_{l\pm}^I(W)=\frac{E+m}{2} \,\int_{-1}^{1} dz \left[ g\, P_l(z) + 
|\vec{q}\,|^2 h \left( P_{l\pm 1}(z)-z P_l(z) \right)\right]~,
\end{equation}
where $z=\cos \theta$, with $\theta$ the c.m. angle between the incoming and the
outgoing pion, $|\vec{q}\,|$ is the modulus of the c.m. three-momentum of any 
of the pions, and $E=\sqrt{m^2+\vec{q}\,^2}$ is the c.m.
nucleon energy. Note that $s$, $t$ and $u$  are the Mandelstam variables 
subject to the constraint $s+t+u = 2m^2 + 2M^2$, with $M$ the pion mass.

\medskip\noindent
As a consequence of unitarity, any partial wave amplitude can also be written 
in terms of a corresponding phase shift $\delta_{l\pm}^I$ as
\begin{equation}
\label{pwf}
T_{l\pm}^I(W)=\frac{8 \pi W}{|\vec{q}\,|} \exp(i\delta_{l\pm}^I)
\sin\delta_{l\pm}^I~.
\end{equation}
This implies the unitarity relation
\begin{equation}
\label{unif}
\hbox{Im}~T_{l\pm}^I(W)=\frac{|\vec{q}\,|}{8 \pi W}|T_{l\pm}^I(W)|^2~,
\end{equation}
which is valid above threshold and before the opening of any inelastic channel.

\medskip\noindent 
The amplitudes $g(W,t)$ and $h(W,t)$ are specially suited for a non-relativistic
treatment of the $\pi$N scattering, as it is carried out e.g. in Heavy Baryon
Chiral Perturbation Theory (HBCHPT) (for a review on HBCHPT see \cite{hbc1}
and an update is given in \cite{hbc2}).\footnote{We remind the reader
on the remarks made in the introduction on the calculation of pion--nucleon scattering in a
relativistic formulation of baryon chiral perturbation theory.} 
However, in a fully relativistic
approach of $\pi$N scattering, it is more convenient to work with the invariant
amplitudes $A(s,t)$ and $B(s,t)$, which are free of kinematical 
singularities and are more suited for studying the analytical properties 
of the scattering amplitude~\cite{Hohler}. In
terms of the invariant functions $A(s,t)$ and $B(s,t)$, the $\pi$N 
scattering amplitudes are written as:
\begin{equation}
\label{AB}
T(W,t)= \bar{u}(p_2,\lambda_2)\left[ A(s,t)+\frac{1}{2}(q_2+q_1)^\mu \gamma_\mu 
 B(s,t) \right] u(p_1,\lambda_1)~,
\end{equation} 
where the $\lambda_i$, appearing in the Dirac spinors, denote the helicities 
of the incoming/outgoing nucleon.
One can derive in a straightforward way the relation between the $A$, $B$ and
$g$, $h$ amplitudes. From the expressions given in ref. \cite{Paul} one obtains, 
\begin{eqnarray}
\label{relation}
g&=&\frac{1}{2(m+E)^2}\left\{ \left(4m(E+m)-t \right)A+
\left[(W+m)(t+4|\vec{q}\,|^2)+4 m w (E+m) \right] B \right\}~, \nonumber \\
h&=&\frac{1}{(m+E)^2}\{(W+m)B-A \}~,
\end{eqnarray}
with $w$ the c.m. pion energy, $w=W-E$.
It is evident from the previous equations that, since the amplitudes $A(s,t)$ 
and $B(s,t)$ are free of kinematical singularities, the same does not hold 
for the amplitudes $g(W,t)$ and $h(W,t)$ due to the dependence on $W=\sqrt{s}$. 
Note that the cut singularities induced by this  dependence have 
nothing to do with the exchange of any particle states. This is the reason why 
they are called kinematical singularities, in contrast to those due to the 
exchange of physical particles, which are called dynamical singularities.
In the invariant amplitudes $A(s,t)$ and $B(s,t)$, only this latter type
of singularity appears.

\medskip\noindent
So far, we have ignored the isospin of the pion ($I_\pi = 1$) and the nucleon 
($I_N = 1/2)$ fields. Consequently, in $\pi$N scattering there are two 
isospin amplitudes, for total isospin $I$=3/2 and $I$=1/2, respectively.
These amplitudes can be written in terms of the $\pi^+ p\rightarrow \pi^+ p$
amplitude, $T_1$, and of the one for $\pi^- p\rightarrow \pi^- p$, $T_2$, as:
\begin{eqnarray}
\label{isospin}
T_{I=3/2}(W,t)&=&T_1(W,t)~, \\ \nonumber
T_{I=1/2}(W,t)&=&\frac{3}{2}T_2(W,t)-\frac{1}{2}T_1(W,t)~.
\end{eqnarray}
If we denote by $A_1(s,t)$ and $B_1(s,t)$ the invariant amplitudes for $T_1(W,t)$
and by $A_2(s,t)$ and $B_2(s,t)$ the corresponding ones for $T_2(s,t)$ then, by
crossing, the following relations arise:
\begin{eqnarray}
\label{crossing}
A_2(s,t) &=& +A_1(u,t)~, \\ \nonumber
B_2(s,t) &=& -B_1(u,t);
\end{eqnarray}
which allow one to consider just the $\pi^+ p\rightarrow \pi^+ p$ reaction in
order to determine the $I=1/2$ and $3/2$ scattering amplitudes. This is similar
to the case of elastic pion--kaon scattering, see e.g. ref.\cite{bkmpik}.

\section{Tree level amplitudes and chiral loops}
\def\theequation{\arabic{section}.\arabic{equation}}
\setcounter{equation}{0}

In this section we construct the basic amplitudes which we will unitarize
in the next section. First, we consider the tree level contributions to $T_1$
arising from the lowest order meson-baryon Chiral Perturbation Theory (CHPT)
Lagrangian, ${\mathcal{L}}^{(1)}_{\pi N}$. Then, the exchange of the delta
isobar $\Delta(1232)$ and of the $N^*(1440)$ and of meson resonances in the 
$t$--channel is also taken into account. Note that
beyond leading order, the chiral Lagrangian contains local contact terms
of second (and higher) order. In our approach, these will be generated by
the explicit contribution from the aforementioned resonances. That this 
is a sensible starting point is based on the observations made in 
ref.\cite{aspects}, where the dimension two low--energy constants were 
fixed from data and it was shown that resonance saturation allows to explain 
these values. Having set up the tree level Lagrangian, 
 we will then consider the ${\mathcal{O}}(p^3)$ HBCHPT 
loop contributions~\cite{mojzis,nadia} to the different isospin 
amplitudes taking the results directly from ref.\cite{nadia}.

\subsection{Pion--nucleon Born terms}
In the two--flavor SU(2) case, the one we are interested in here, the lowest 
order CHPT pion--nucleon Lagrangian takes the form:
\begin{equation}
{\mathcal{L}}^{(1)}_{\pi N}
=\bar{\Psi}(i\gamma_\mu D^\mu-m+\frac{g_A}{2}\gamma^\mu
\gamma_5 u_\mu)\Psi
\end{equation}
In this expression one has to take into account that the nucleon mass,
$m$, and the axial-vector coupling $g_A$ are to be taken
in the chiral limit ($m_u = m_d = 0, m_s$ fixed). However, in
the following we will use the physical values of these quantities since, 
when including the chiral loops at ${\mathcal{O}}(p^3)$~\cite{nadia}, they will be 
renormalized, up to this order, to their physical values. In the following we 
will take $g_A=1.26$ and $m=938.27$ MeV. The choice
of the proton mass is justified because we are only considering 
the scattering of pions off the proton.  From this process 
we derive the two isospin amplitudes $\pi$N scattering, as 
explained above. On the other hand, the proton $p$ and the neutron $n$ 
fields are collected in the isospinor $\Psi$
\be
\label{isos}
\Psi=\left(
\begin{array}{c}
p\\
n
\end{array}
\right) ,
\ee
the covariant derivative $D_\mu$ acting on the nucleon field is given by
(note that we do not consider external fields and isospin breaking here) 
\ba
\label{cov}
D_\mu \Psi&=&\partial_\mu \Psi+\Gamma_\mu \Psi~, \nonumber \\
\Gamma_\mu&=&\frac{1}{2}[u^\dagger,\partial_\mu u]~.
\ea
Here, $u_\mu=i \,u^\dagger\partial_\mu U u^\dagger$ and
$\displaystyle{U=u^2=\exp[i\tau^j\pi^j/F_\pi]}$, where $\tau^j$ refers 
to the Pauli matrices (in isospin space).
The tree level diagrams from ${\mathcal{L}}^{(1)}_{\pi N}$ are depicted in fig.1.
The corresponding expressions are:
\begin{eqnarray}
\label{l1b}
        T_1^a&=& \frac{g_A^2}{4\,F_\pi^2}\left(4m+ \barr{R} 
\frac{u+3 m^2}{u-m^2} \right)~,\\ \nonumber
        T_1^b&=& -\frac{\barr{R}}{4\, F_\pi^2}~, 
\end{eqnarray}
with $F_\pi=92.4$ MeV the (weak) decay constant of the pion and 
$R_\mu=(q_1+q_2)_\mu$.  Note that the direct
nucleon exchange diagram does not contribute to $\pi^+ p\rightarrow \pi^+ p$.

\begin{figure}[ht]
\centerline{\includegraphics[width=0.2\textwidth,angle=-90]{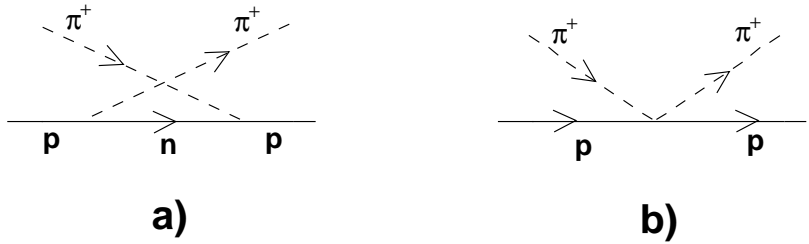}}
\caption{Diagrams from ${\mathcal{L}}^{(1)}_{\pi N}$: 1.a) Crossed nucleon 
exchange, 1.b) Seagull.}
\end{figure}

\subsection{Baryon resonances}
Next, we consider resonance excitations in the $s$-- (or $u$--)channel. 
We treat explicitly the spin--3/2 $\Delta (1232)$ and also the
first even--parity excited state of the nucleon, the Roper $N^*
(1440)$. Most important for the energy--range we are considering is the
$\Delta(1232)$ resonance. It contributes to $\pi$N scattering through
the $\Delta(1232)$ exchange diagrams, see fig.2. The
interactions of the $\Delta$ field with pions and nucleons are given by the
following effective Lagrangian, which is the leading term of an appropriate chiral
invariant interaction~\cite{hbc1}
\begin{equation}
\label{delta}
{\mathcal{L}}_{\Delta \pi N}=\frac{g_{\Delta \pi N}}{M} \bar{\Delta}^\nu 
\vec{T}^\dagger \Theta_{\nu \lambda}(Z) N \partial^\lambda \vec{\pi}+
{\rm h.c.}~,
\end{equation}
where $g_{\Delta \pi N}$ is the $\Delta \pi N$ coupling and $M=139.57$ is the 
charged pion mass. This choice is justified since we are considering 
elastic $\pi^+ p$ scattering. On the other hand, $\vec{T}$ is the 
$\frac{1}{2}\rightarrow \frac{3}{2}$ isospin transition operator which 
satisfies 
\begin{equation}
\label{sum}
\sum_{\lambda_\Delta}
T_b|\frac{3}{2}\lambda_\Delta\rangle \langle
\frac{3}{2}\lambda_\Delta|T_a^\dagger=\delta_{ab}
-\frac{1}{3}\tau_b\tau_a~.
\end{equation}
Furthermore, the Dirac matrix operator $\Theta_{\mu\nu}(Z)$ is given as
\begin{equation}
\label{theta}
\Theta_{\mu \nu}(Z)=g_{\mu \nu}+\left(\frac{1}{2}(1+4 Z)A+Z \right)\gamma_\mu 
\gamma_\nu=g_{\mu \nu}-(Z+\frac{1}{2})\gamma_\mu \gamma_\nu~,
\end{equation} 
with the right-hand side valid for $A=-1$. The quantity $Z$ is called
an off-shell parameter, which enters the fully relativistic $\Delta \pi N$ vertices.
Finally, the spin--3/2 propagator (momentum $P$, $\nu$ to $\mu$) is:
\begin{equation}
\label{32pro}
-i
\frac{\barr{P}+m_\Delta}{P^2-m_\Delta^2}\left(g_{\mu\nu}-\frac{1}{3}\gamma_\mu
\gamma_\nu-\frac{2 P_\mu P_\nu}{3 m_\Delta^2}+\frac{P_\mu \gamma_\nu-
P_\nu \gamma_\mu}{3 m_\Delta} \right)~,
\end{equation}
with $m_\Delta$ the mass of the delta resonance, we take $m_\Delta=1232$ MeV. 
Note that this mass used in the tree graphs is not necessarily equal to
the real part of the S--matrix pole. We will come back to this point
later on.
For the diagram represented in fig.2a, corresponding to the direct exchange of
the $\Delta$, we find

\begin{figure}[ht]
\centerline{\includegraphics[width=0.2\textwidth,angle=-90]{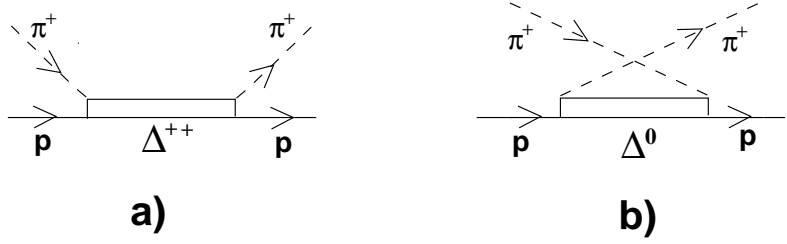}}
\caption{Diagrams with the exchange of the $\Delta(1232)$ resonance.}
\end{figure}

\begin{eqnarray}
\label{deltaam}
T_1^{\Delta .a}&=&A_1^{\Delta .a}+\frac{1}{2}\barr{R} B_1^{\Delta .a}~, 
 \nonumber \\
A_1^{\Delta .a}&=&
{\frac{{{{g_{\Delta \pi N}}}^2}}{6\,{{{m_{\Delta}}}^2}\,{{{M}}^2}\,
(s-m_\Delta^2)}}\,\left\{ 
{{{m_{\Delta}}}^2}\,{m}\,
        \left[ 6\,{{{M}}^2} - 2\,s - 3\,t - 4\,s\,{\kappa^2} + 
          {{{m}}^2}\,\left( 2 + 4\,{\kappa^2} \right)  \right]\right . 
           \nonumber \\ &+& 
       {{m_{\Delta}}^3}\,\left[ 6\,{{{M}}^2} - 2\,s - 3\,t +
       4\,s\,\kappa - 8\,s\,{\kappa^2} + 
          2\,{{{m}}^2}\,\left( 1 - 2\,\kappa + 4\,{\kappa^2}
           \right)  \right] \nonumber \\  &+& 
       2\,{m_{\Delta}}\,\left[ -{{{M}}^4} - {{{M}}^2}\,s
       - 2\,{s^2}\,\kappa + 
          4\,{s^2}\,{\kappa^2}  
          + {{{m}}^2}\,
           \left( {{{M}}^2} + 2\,s\,\kappa 
           - 4\,s\,{\kappa^2} \right)  \right] \nonumber \\ 
           &-& \left .
       {m}\,\left[ {{{m}}^4} + {{{M}}^4} + 
          2\,{{{M}}^2}\,s + {s^2} - 4\,{s^2}\,{\kappa^2} - 
          2\,{{{m}}^2}\,\left( {{{M}}^2} + s - 2\,s\,{\kappa^2}
          \right)  \right]  \right\}~, \nonumber \\ 
B_1^{\Delta .a}&=&
{\frac{{{{g_{\Delta \pi N}}}^2}}{6\,{{{m_{\Delta}}}^2}\,{{{M}}^2}
(s-m_\Delta^2)}}\,
\left\{ -{{{m}}^4} - 
       {{\left( {{{M}}^2} + s - 2\,s\,\kappa \right) }^2} 
       + 2\,{{{m}}^2}\,\left( {{{M}}^2} + s - 2\,s\,\kappa -
       2\,s\,{\kappa^2} \right) \right . \nonumber \\  &+& 
       4\,{{{m_{\Delta}}}^3}\,{m}\,\left( 1 - 2\,\kappa + 
 4\,{\kappa^2}\right) +
       2\,{m_{\Delta}}\,{m}\,
        \left( {{{m}}^2} - {{{M}}^2} - s + 4\,s\,\kappa -
        8\,s\,{\kappa^2} \right) \nonumber \\  
        &+& \left . 
       {{{m_{\Delta}}}^2} \left[ 4\,{{{M}}^2} - 3\,t -
       4\,{{{M}}^2}\,\kappa - 
          4\,s\,\kappa + 4\,s\,{\kappa^2} + 
          4\,{{{m}}^2}\,\left( 1 + \kappa + {\kappa^2}
          \right)  \right] \right\}~, 
\end{eqnarray}
with $\kappa=Z+\frac{1}{2}$.
\noindent
The diagram of fig.2b can we obtained from the former one just by crossing. The
result is:
\begin{eqnarray}
\label{td2}
A_1^{\Delta.b}(s,t)&=&+\frac{1}{3} A_1^{\Delta.a}(u,t)~, \nonumber \\
B_1^{\Delta.b}(s,t)&=&-\frac{1}{3} B_1^{\Delta.a}(u,t)~.
\end{eqnarray}

\medskip \noindent
Let us consider now the exchange of the $N^*(1440)$ Roper resonance, cf. fig.3. 
Although this resonance is heavier than the $\Delta(1232)$ and not as
strongly coupled to the $\pi$N system, in  ref.\cite{bora} it was shown
that the Roper octet can have an appreciable interference with the contributions
coming from the exchange of scalar resonances. The Lagrangian for the coupling
$N^* N \pi$ is~\cite{cucu}
\be
\label{roper}
{\mathcal{L}}_{N^*N\pi}=\frac{g_A}{4}\sqrt{R}\,\bar{\Psi}_{N^*}
\gamma_\mu \gamma_5 u^\mu \Psi_N + {\rm h.c.}
\ee 
with $\sqrt{R}=0.53\pm0.04$~\cite{cucu} from the width of the $N^*(1440)$ 
determined in ref.\cite{hoh} (note that the often used Breit--Wigner
fits to this resonance lead to an appreciably larger width). 
The resulting amplitude is given by:
\be
\label{trop}
T_1^{N^*}=\frac{g_A^2 R}{8F_\pi^2(u-m_R^2)}\left\{(u-m^2)(m+m_R)+
\frac{\barr{R}}{2}(u+m^2+2 m\, m_R)\right\}
\ee
where $m_R=1440$ MeV is the mass of the Roper resonance.

\begin{figure}[ht]
\centerline{\includegraphics[width=0.2\textwidth,angle=-90]{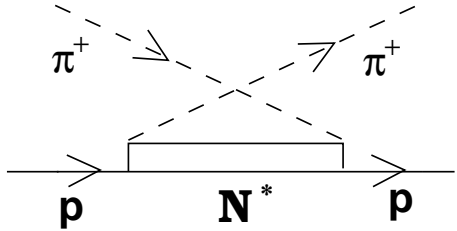}}
\caption{s--channel exchange of the Roper resonance.}
\end{figure}

\subsection{Meson resonances}
We now turn to the $t$--channel exchange of meson resonances.
First, we consider the exchange of an isosinglet scalar resonance, cf. fig.4. 
This state should account for both the contribution of the lightest scalar 
singlet and octet $I=0$ states, which in large $N_c$ (where $N_c$ denotes the
number of colors) should be degenerate. The 
SU(2) $S\pi\pi$ interaction can be written as~\cite{aspects}
\be
\label{sla}
{\mathcal{L}}_{S\pi\pi}=S\left[\bar{c}_m\,{\rm Tr}(\chi_+)+\bar{c}_d\,
{\rm Tr}(u_\mu u^\mu)\right]
\ee 
with $\chi_+ = {\rm diag}(M^2,M^2)$ and the $\bar{c}_d$, $\bar{c}_m$
are two lowest order coupling constants. 
In ref.\cite{nd}, where a simultaneous fit to the all SU(3) related S--wave 
meson--meson  scalar channels was done combining CHPT with the exchanges of
resonances in the $s$--channel via the $N/D$ method,
the scalar singlet appears with a mass around 1~GeV and 
the octet is higher in energy, around 1.35~GeV. 
In ref. \cite{jamin} it was shown that 
the requirement that the tree level $K\pi$, $K\eta$ and $K\eta'$ scalar
form factors vanish at infinity fixes these coupling in terms
of the pion decay constant, 
\be
\label{coupling}
\bar{c}_d=\bar{c}_m=\frac{F_\pi}{\sqrt{6}}.
\ee
Requiring also the saturation of the ${\mathcal{O}}(p^4)$ CHPT counterterms for
the former couplings, the nonet mass is then fixed to be around 1.2~GeV, which
lies in an intermediate value between the singlet and octet mass given in
ref.\cite{nd}. The well--known 
$\sigma(550)$ meson, see e.g. refs.\cite{npa,nd,juelich} should be
considered as a reflection of 
the strong S--wave $\pi\pi$ interactions, loop physics and hence subleading in 
large $N_c$ counting rules (this has already been stressed a long time
ago, see~\cite{ulfsig}). 
Other QCD inspired approaches, which also 
establish that the lightest preexisting scalar meson should be around 1~GeV, 
are collected in refs.\cite{barcelona,thomas}. While the contributions of a 
preexisting scalar resonance appear at ${\mathcal{O}}(p^2)$ in the chiral counting, 
the crossed pions loops, and hence their associated $\sigma$ pole, begins at 
${\mathcal{O}}(p^3)$.
The coupling of the scalar $I=0$ resonance to the nucleon is calculated easily 
from the Lagrangian
\be
\label{snn}
{\mathcal{L}}_{S\,NN}=-g_s\bar{\Psi}\Psi~.
\ee
{}From eqs.(\ref{sla}),(\ref{snn}) we obtain the scalar meson exchange
contribution to $\pi$N scattering,
\be
\label{st}
T_1^S=-2\,g_S\,\frac{\bar{c}_m\,2M^2+\bar{c}_d\,(t-2M^2)}{F_\pi^2\,(t-M_S^2)}
\ee
with $M_S$ the mass of the scalar resonance.

\begin{figure}[ht]
\centerline{\includegraphics[width=0.2\textwidth,angle=-90]{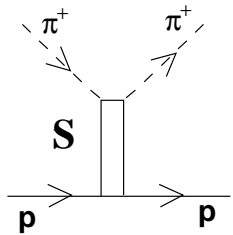}}
\caption{t-channel exchange of an isosinglet resonance, $S$.}
\end{figure}

\medskip\noindent
Next, we consider the exchange of the $\rho(770)$ meson in the $t$--channel, 
fig.5. Following refs.\cite{ecker,ulf}, we consider the tensor representation
for the vector fields. In this way, we generate terms in the chiral
expansion of the pion--nucleon scattering amplitude that
begin at ${\mathcal{O}}(p^2)$. On the other hand, if we were to use the vector
representation, the contributions from the exchange of the $\rho$ would begin to
appear only for orders higher than ${\mathcal{O}}(p^2)$. Consequently,  extra
local contact terms would  be needed in order to make both
representations, vector  and tensor, equivalent~\cite{leut,ulf}. 
The tensor field representation is most useful in constructing chiral
invariant building blocks for spin--1 mesons coupled to nucleons,
pions and external fields (like e.g. photons).
The $\rho \pi^+ \pi^-$ vertex can be obtained from the Lagrangian~\cite{ecker}
\be
\label{rpla}
{\mathcal{L}}_{V\phi\phi}=i\frac{G_V}{\sqrt{2}} 
{\rm Tr}\left[W_{\mu\nu} u^\mu u^\nu\right]~,
\ee
where $W_{\mu\nu}$ refers to the tensor field representing the octet
of the lowest--lying vector resonances,
\be
W_{\mu\nu} = \left(
  \begin{array}{ccc}
\frac{\rho^0}{\sqrt{2}}+\frac{\omega_8}{\sqrt{6}} & \rho^+ & K^{*\,+} \\
 \rho^-&-\frac{\rho^0}{\sqrt{2}}+\frac{\omega_8}{\sqrt{6}}& K^{*\,0}\\
 K^{*\,-}& \bar{K}^{*\,0}&-\frac{2\omega_8}{\sqrt{6}}
  \end{array}
\right)_{\mu\nu}
\ee
such that 
\be
\langle 0|W_{\mu\nu}|V,p \rangle =
i M_V^{-1}[p_\mu\epsilon_\nu(p)-p_\nu\epsilon_\mu(p)]~,
\ee
with $\epsilon_\mu(p)$ the usual polarization vector for the vector state
$|V,p\rangle$ with a mass in the chiral limit of $M_V\approx M_\rho=770$~MeV~\cite{ecker}. 
The propagator is given by
\be
\frac{M_V^{-2}}{M_V^2-p^2-i\epsilon}\left[g_{\mu\rho}g_{\nu\sigma}(M_V^2-p^2)+
g_{\mu\rho}p_\nu p_\sigma-g_{\mu\sigma}p_\nu p_\rho-(\mu \leftrightarrow \nu)
\right].
\ee
The coupling $G_V$ can be obtained from the decay 
$\rho \to 2 \pi$
and one obtains $G_V=69\,$MeV. Another possible way of determining
$G_V$ is from fitting the  charge radius of the pion. That leads to a
reduced value of $G_V=53\,$MeV~\cite{gl}. We will distinguish below 
between both values when we
present our results. One has to take into account that in ref.~\cite{ecker,ulf}
an SU(3) notation is considered. We will also use this three--flavor notation
to work out the $\rho$--couplings. The only
difference for our considerations is that the Pauli matrices, $\tau^i$,
are changed to the Gell-Mann matrices $\lambda^i$. In this way the
SU(3) representation of the Goldstone boson octet reads
\be
U=\exp \biggl( {i\frac{\lambda^j \phi^j}{F_\pi}}\biggr) =\left(
  \begin{array}{ccc}
\frac{\pi^0}{\sqrt{2}}+\frac{\eta_8}{\sqrt{6}} & \pi^+ & K^{+} \\
 \pi^-&-\frac{\pi^0}{\sqrt{2}}+\frac{\eta_8}{\sqrt{6}}& K^{0}\\
 K^{-}& \bar{K}^{0}&-\frac{2\eta_8}{\sqrt{6}}
  \end{array}
\right)~.
\ee

\begin{figure}[ht]
\centerline{\includegraphics[width=0.2\textwidth,angle=-90]{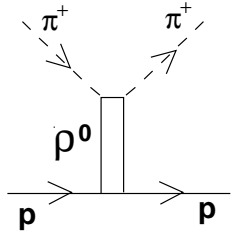}}
\caption{t-channel exchange of the $\rho(770)$ resonance.}
\end{figure}

\noindent
On the other hand, the  $\rho NN$ vertex can be inferred from
ref.\cite{ulf}  from the general vector to baryon octet coupling via
\ba
\label{rnla}
{\mathcal{L}}_{VNN}&=&R_{D/F} Tr[\bar{B} \sigma^{\mu\nu}(W_{\mu\nu},B)] 
+ S_{D/F} Tr[\bar{B} \gamma^\mu(\partial^\nu W_{\mu\nu},B)] \nonumber \\
&+& T_{D/F} Tr[\bar{B} \gamma^\mu(\partial_\lambda W_{\mu\nu},
\partial^\lambda \partial^\nu B)]
+ U_{D/F} Tr[\bar{B}\sigma^{\lambda\nu}(W_{\mu\nu},\partial_\lambda 
\partial^\mu B)]~.
\ea 
Here, the symbol $(A,B)$ denotes either the commutator, $(A,B)=[A,B]$, or the
anticommutator, $(A,B)=\{A,B\}$. In the first case, the coupling constant has
the subscript $F$, in the second one the subscript $D$ applies. For
instance, the first term in the previous equation thus means
\be
\label{exemple}
R_D Tr[\bar{B}\sigma^{\mu\nu}\{W_{\mu\nu},B\}]+R_F
Tr[\bar{B}\sigma^{\mu\nu}[W_{\mu\nu},B]]~.
\ee
{}From the Lagrangians given in eqs.(\ref{rpla}) and (\ref{rnla}), one obtains 
for the diagram depicted in fig.5:
\be
\label{rho}
T_1^\rho=\frac{2 G_V \sqrt{2}}{F_\pi^2(M_\rho^2-t)}\left\{R_+\frac{u-s}{2}+
\frac{\barr{R}}{2}\left[2 m R_+ +\frac{t}{2}(S_+-\frac{m\,U_+}{2})-
\frac{t^2\,T_+}{8}\right]\right\}~,
\ee
with $R_+=R_D+R_F$ and analogously for $S_+$, $U_+$ and $T_+$. All the 
former unknown parameters, except $T_+$, can be recast in terms of more 
familiar ones, by comparing eq.(\ref{rho}) with 
the result that one would obtain by making use of the standard vector
representation for the 
$\rho$ meson. In this case, instead of the Lagrangian given in eq.(\ref{rnla}), 
one has:
\be
\label{rnvla}
{\mathcal{L}}_{\rho\,NN}=\frac{1}{2} \, g_{\rho NN} \,\bar{\Psi}
\left[\gamma^\mu \vec{\rho}_\mu\vec{\tau}-
\frac{\kappa_\rho}{2\,m}\sigma^{\mu\nu}\partial_\nu\vec{\rho_\mu}\vec{\tau}
\right]\Psi~,
\ee
with $g_{\rho NN}$ the vector coupling of the $\rho$ to two nucleons and
$\kappa_\rho$ the ratio between the tensor and the vector couplings.
After these preliminaries, the final expression that we obtain for the 
$\rho$-exchange contribution is given by:
\ba
T_1^\rho&=&\frac{t\,G_V\,g_{\rho NN}}{M_\rho F_\pi^2(t-M_\rho^2)}\left(
\frac{1+\kappa_\rho}{2}\barr{R}+\frac{u-s}{4\,m}\kappa_\rho \right) \nonumber
\\ &-&\frac{g_{\rho NN}\,\kappa_\rho\,G_V}{F_\pi^2\,M_\rho}
\left(\frac{\barr{R}}{2}+\frac{u-s}{4\,m}\right)+
\frac{t^2\,\sqrt{2}\,G_V}{4\,F_\pi^2\,(t-M_\rho^2)}\,T_+~.
\ea
Note that the coupling $\sim t^2 \, T_+$ has not been considered
before although it is solely based on chiral symmetry. 
Its relevance for the $\rho$NN phenomenology will be discussed below.

\subsection{Chiral pion loops}
As explained in detail in the introduction, we consider here the
${\cal O}(p^3)$ one--loop contributions obtained in the framework of
heavy baryon chiral perturbation theory.
After renormalizing all the bare parameters, these loop contributions
are given in ref.~\cite{nadia}.
Their contributions to the spin--non--flip and spin--flip
amplitudes with given total isospin ($I=3/2$ or $1/2$), 
$g_{3/2}$, $g_{1/2}$ and $h_{3/2}$, $h_{1/2}$ are:
\ba
\label{gn}
g_{3/2\,{\rm loop}}&=&g^+_{{\rm loop}}-g^-_{{\rm loop}} \nonumber \\
g_{1/2\,{\rm loop}}&=&g^+_{{\rm loop}}+2\,g^-_{{\rm loop}}
\ea
where we have followed the notation of ref.~\cite{nadia} to denote their
calculated loop functions. For the $h_{3/2\,{\rm loop}}$ and
$h_{1/2\,{\rm loop}}$ functions
analogous relations with the functions $h^+_{{\rm loop}}$, $h^-_{{\rm
    loop}}$ as the ones 
shown in eq.(\ref{gn}) hold.

\section{Unitarization}
\def\theequation{\arabic{section}.\arabic{equation}}
\setcounter{equation}{0}

As a consequence of eqs.(\ref{pw}) and (\ref{relation}), any $\pi$N partial 
wave amplitude has besides the dynamical cuts, which are due to the exchange of particle
states in the
$s$--, $t$-- or $u$-- channel, further kinematical singularities
 arising from the $\sqrt{s}$--dependence in eq.(\ref{relation}). In the
 following, we will consider a partial wave amplitude 
as a function of $W$, instead of $s$, and hence the kinematical singularities 
disappear. 
The dynamical cuts are  usually divided into  two classes: the physical
or unitarity cut and the unphysical cuts. The physical cut is required by 
unitarity, eq.(\ref{unif}), and the unphysical cuts are due to the exchanges 
of particles in the crossed channels. A more detailed discussion of
the ideas underlying this approach are given in ref.\cite{nd}. Here,
we are concerned with the extension of such a scheme to the
pion--nucleon sector.

\subsection{The right--hand cut}
We consider first the physical or right--hand cut and neglect, for 
the moment, the 
unphysical ones. In order to obtain a unitarized $\pi$N amplitude, we follow 
ref.\cite{nd}, in which meson--meson scattering was considered. In
this work, making use of the $N/D$ method~\cite{chew}, the 
most general structure of a meson--meson partial wave amplitude, when no 
unphysical cuts are considered, is given. If we denote by $T_l(s)$ a meson--meson
partial wave with orbital angular momentum $l$ the central result of
ref.\cite{nd} takes the form:
\ba
\label{mes}
T_l(s)&=&\frac{N(s)}{D(s)} \nonumber \\
N(s)&=&1 \nonumber \\
D(s)&=&\sum_i \frac{R_i}{s-s_i}+a(s_0)-\frac{s-s_0}{\pi}\int_{s_{\rm thr}}
^\infty ds' \frac{\rho(s')}{(s'-s)(s'-s_0)}~.
\ea
Here $\displaystyle{\rho(s)=
{|\vec{q}\,|}/(8\,\pi\,\sqrt{s})}$, see eq.~(\ref{unif}),
 $s_{\rm thr}$ is the
value of $s$ for the threshold of the reaction, $s_0$ is the subtraction point,
$a(s_0)$ a subtraction constant and each pole of the sum is referred
to as a CDD pole after
ref.\cite{castillejo}, with $s_i$ the pole position and $R_i$ the corresponding
residue. Note that a CDD pole is a pole of the $D(s)$ function so that it
corresponds to a zero of the partial wave $T_l(s)$.

\noindent
In order to translate the result of eq.(\ref{mes}) to the $\pi$N case, we
first consider how the right--hand cut, given by the integral in eq.(\ref{mes}),
looks like in the complex $W$--plane. Note that because $W=\sqrt{s}$ and
since the square root is a double--valued function, we will have two cuts in the 
$W$--plane. One of these cuts extends from
$W>W_{\rm thr}=M+m$ to $+\infty$ and the other one from $-\infty$ to
$-W_{\rm thr}$.  Therefore, the 
contour of integration in the complex $s$--plane used in ref.\cite{nd} 
to obtain eq.(\ref{mes}), Fig.6, transforms in the $W$-plane to the
one shown in fig.7. In this last figure, each line is denoted by two
numbers $i,j$: $i$ refers to the corresponding line in Fig.6 and $j=1$ means 
first
sheet for $W=\sqrt{s}$ (Im$W>0$) and $j=2$ refers to the second one (Im$W<$0).
Then the dispersion integral in the $W$-plane for $D(W)$, with a subtraction 
in $W_0$, will be:
\begin{figure}[ht]
\centerline{\includegraphics[width=0.3\textwidth,angle=-90]{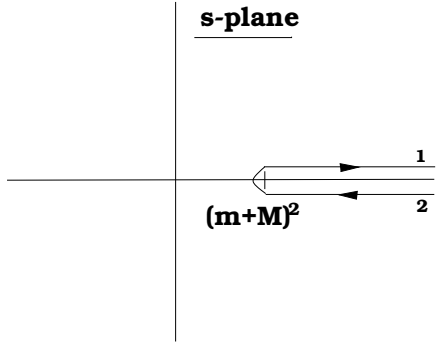}}
\caption{Contour of integration for the Cauchy representation of $D(s)$ 
eq. (\ref{mes}). The enclosing circle extending to infinity is not shown.}
\end{figure}
\ba
\label{wsi}
&&\frac{W-W_0}{2\pi i}\left\{ \int_{W_{\rm thr}}^\infty dW' 
\frac{D(W')_{11}}{(W'-W)(W'-W_0)}
-\int_{W_{\rm thr}}^\infty dW' 
\frac{D(W')_{22}}{(W'-W)(W'-W_0)} \right .\nonumber \\
&+& \left . \int_{W_{\rm thr}}^\infty dW' 
\frac{D(-W')_{21}}{(W'+W)(W'+W_0)}
-\int_{W_{\rm thr}}^\infty dW'
\frac{D(-W')_{12}}{(W'+W_0)(W'+W)} 
\right\}
\ea
where we have indicated by a subindex $ij$ the line in Fig.7 over which the
integral has to be taken. Making use in the former expression of the fact that
$T(W)=T^*(W^*)$, where $^*$ means complex conjugation, we then have
\be
\label{wsi2}
\frac{W-W_0}{\pi}\left\{\int_{W_{\rm thr}}^\infty dW' \left[
\frac{\hbox{Im}D(W')_{11}}{(W'-W)(W'-W_0)}+
\frac{\hbox{Im}D(-W')_{21}}{(W'+W)(W'+W_0)} \right] \right\}
\ee
Since the line denoted by 21 in Fig.7 corresponds to the line 2 in fig.6, then
\be
\hbox{Im}D(-W')_{21}=-\hbox{Im}D(W')_{11}=\rho(W')
\ee
and hence we finally have
\ba
\label{wsi3}
&&\frac{W-W_0}{\pi}\int_{W_{\rm thr}}^\infty dW' \hbox{Im}D(W')_{11}\left[
\frac{1}{(W'-W)(W'-W_0)}-\frac{1}{(W'+W)(W'+W_0)}\right] \nonumber \\
&=&-\frac{s-s_0}{\pi}
\int_{s_{\rm thr}}^\infty ds'\frac{\rho(s')}{(s'-s)(s'-s_0)}~,
\ea
with $s_0=W_0^2$ and $s'=W^{'\,2}$. Thus, we recover the same integral
as in eq.(\ref{mes}) but this time in the complex $W$--plane. The
remaining terms in eq.(\ref{mes}) are not altered: the subtraction
constant is, of course, also present in 
the $W$--variable case, and the CDD poles will be now in
the form $\displaystyle{R_i'}/({W-W_i})$. Note that, as stated above, these
poles represent just zeros of the $T_{l\pm}^I(W)$ partial wave and then have to
appear as poles of the $D(W)$ function but now in the $W$--variable. Note that a
complex CDD pole in the $s$--variable gives rise to four poles in the
$W$--variable. Hence, instead of eq.(\ref{mes}) we will have for any $\pi$N
partial wave $T_{l\pm}^I(W)$:
\be
\label{nsi}
T_{l\pm}^I(W)=\left[\sum_i \frac{R_i'}{W-W_i}+a(s_0)-\frac{s-s_0}{\pi}
\int_{s_{\rm thr}}^\infty ds'\frac{\rho(s')}{(s'-s)(s'-s_0)} \right]^{-1}
\ee
It is important to stress that in what follows we will fix the
subtraction constant at a point $s_0$ where it is real. This, however,
is done for convenience and need not be done. For $a(s_0)$ to be real,
we have $s_0 \le s_{\rm thr}$, with $ s_{\rm thr}$ the threshold of
the reaction under consideration. Note also that eq.(\ref{nsi}) allows
to accomodate the possible pole of $T_{l\pm}^I (W)$ at $W=0$, due to
the kinematical singularities~\cite{Hohler}, which give rise to a
$1/W$ factor in our definition of the partial wave amplitudes.
\begin{figure}[ht]
\centerline{\includegraphics[width=0.3\textwidth,angle=-90]{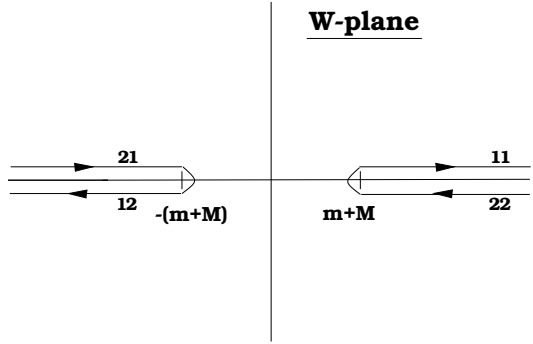}}
\caption{Contour of integration for the Cauchy representation of $D(W)$ 
eq. (\ref{nsi}). The enclosing circle extending to infinity is not shown.}
\end{figure}

\medskip\noindent 
A connection with the large $N_c$ limit~\cite{TW} is very enlighting. In this
limit of QCD the dominant terms of any meson-baryon amplitude are of order
${\mathcal{O}}(N_c^{0})$ and
given by the local terms and pole contributions. 
These are obtained in eq.(\ref{nsi}) by the  inverse
of the CDD poles and the leading piece of the subtraction constant,
which are of order 
${\mathcal{O}}(N_c^0)$. Note that the integral in this equation, which contains
the 
unitarity loops in the $s$--channel, is of order ${\mathcal{O}}(N_c^{-1})$ and hence is
subleading in 
the large $N_c$ counting, as expected. As a consequence, we
split the subtraction constant in two pieces:
\be
\label{subc}
a(s_0)=a^L+a^{SL}(s_0)
\ee
where $a^L$ is ${\mathcal{O}}(N_c^0)$ and independent of the subtraction point
$s_0$, while $a^{SL}(s_0)$ is ${\mathcal{O}}(N_c^{-1})$ and is linked to the
integral so that their sum is independent of the subtraction point
$s_0$.  Therefore, we introduce the function $g(s)$
\be
\label{g(s)}
g(s)=a^{SL}(s_0)-\frac{1}{\pi}\int_{s_{\rm thr}}^\infty ds' 
\frac{\rho(s')}{(s'-s)(s'-s_0)}~.
\ee
On the other hand, we also need the $K$ function
\be
\label{kf}
K(W)=\left[\sum_i \frac{R_i'}{W-W_i'}+a^L\right]^{-1}~.
\ee
Hence, in the absence of unphysical cuts, we can write 
\be
\label{tnuc}
T_{l\pm}^I=\left[K(W)^{-1}+g(s)\right]^{-1}=\frac{K(W)}{1+K(W)\,g(s)}~,
\ee
in close analogy with a $K$-matrix formalism.
In this derivation we have considered the isospin limit. 
The extension of  this is framework to include strong isospin violation
is based on a coupled channel formalism as detailed in ref.~\cite{nd}.
Further isospin breaking due to electromagnetic effects is harder to
include and will not be considered in what follows.

\subsection{The left--hand cut}
We have discussed in detail the case of no unphysical cuts since we will take
this case as our starting point. The unphysical cuts will be treated in
a perturbative way. The perturbation will be done in a chiral loop 
expansion of the
unphysical cuts. In this way, our final calculation will include their
contributions up to one loop calculated at ${\mathcal{O}}(p^3)$ in 
HBCHPT.\footnote{Strictly speaking the HBCHPT counting is referred to the $g(W,t)$ and
$h(W,t)$ functions because the coefficient $(E+m)/2$ in
front of eq.(\ref{pw}) is not expanded. This is based on the fact that for a proper
wavefunction renormalization and matching to the relativistic theory, one should
keep this prefactor as it is given here. For a discussion on these
issues, see ref.\cite{FMSZ}.} This way of proceeding has already been used in 
ref.\cite{WW} to study the strong
$W_L W_L$ scattering in the strongly interacting Higgs sector and is also  being
used in ref.\cite{jamin} to study the $K\pi$, $K\eta'$ S-wave scattering. 
 
\medskip\noindent
Let us now discuss in detail our treatment of the left--hand cuts.
Denote by $T_{{\rm tree+loop}}^{I\,l\pm}$ any $\pi$N partial wave amplitude
resulting from the contributions discussed in sect.~3. If we assume the
saturation of the HBCHPT counterterms up to ${\mathcal{O}}(p^3)$ by resonance 
exchange and perform a non--relativistic expansion of 
$T_{{\rm tree+loop}}^{I\,l\pm}$, we would recover the HBCHPT amplitude up to 
${\mathcal{O}}(p^3)$ given in ref.\cite{nadia}.
We now write
\be
\label{ndtl}
T_{l\pm}^I(W)=\frac{N(W)}{D(W)}=\frac{N(W)}{1+N(W)\,g(s)}
\ee
with $N(W)=K(W)(1+\delta N(W))$. For the case of not including any unphysical contributions,
i.e. $\delta N(W)=0$, we recover eq.(\ref{tnuc}). We expand this representation
of the partial wave amplitudes in order to reproduce
the one loop ${\mathcal{O}}(p^3)$ HBCHPT result
$T_{{\rm tree+loop}}^{I\,l\pm}$:
\be
\label{en}
T_{{\rm tree+loop}}^{I\,l\pm}(W)=N(W)-N_1(W)^2\,g(s)~,
\ee
where $N_1(W)$ is the $\Opu$ contribution of $N(W)$ and is given by the lowest 
order HBCHPT amplitudes. Consequently
\be
\label{new}
N(W)=T_{{\rm tree+loop}}^{I\,l\pm}(W)+N_1(W)^2\,g(s)~.
\ee
We note that $T_{\rm tree+loop}^{I\,l\pm}$ satisfies perturbative unitarity up to $\Opt$
in the HBCHPT chiral counting, which can be traced back to the proper
wave function renormalization of the tree graphs. However, to assure
exact unitarity also for the loop graphs (which is of higher order in 
HBCHPT), we have to multiply $g(s)$ in the previous equation by a correction factor,
$\displaystyle{\varrho=(2\,W)/(E+m)=1+\Opd}$. This assures a complete
matching of the heavy baryon amplitude to the relativistic one with
respect to the constraints from unitarity. Stated differently, 
if would not multiply the function $g(s)$ by $\varrho$, then the exact unitarity 
would be lost.
Note that from the last equation and in eq.(\ref{ndtl}), the $N(W)$ and the 
$D(W)$ functions satisfy the $N/D$ equations up
to the same order than our input $T_{{\rm tree+loop}}^{I\,l\pm}(W)$, that is, up to 
one loop calculated at $\Opt$ in HBCHPT. On the other hand, the $D$ function
satisfies them exactly for the right--hand cut. As a result, our final expression 
for any partial wave, eqs.(\ref{ndtl}) and (\ref{new}), satisfies unitarity to
all  orders. We note that the low--energy representation could be
further improved by matching to the fourth order HBCHPT amplitude,
which should be available in the near future.

\begin{figure}[ht]
\centerline{\includegraphics[width=0.3\textwidth,angle=0]{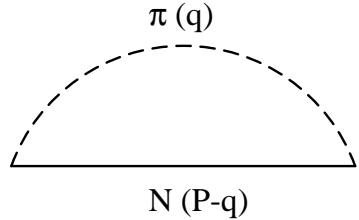}}
\caption{Loop graph  corresponding to the function $g(s)$.}
\end{figure}

\noindent
Before finishing this section, we give an alternative expression for $g(s)$ more
suited than eq.(\ref{g(s)}) for a direct comparison with HBCHPT. This new
expression is derived making use of dimensional regularization. Since the
function $g(s)$  corresponds, up to a constant, to the loop graph with one pion and one
nucleon propagator, see fig.8, we can write:
\ba
\label{gs2}
g(s)&=&\bar{a}^{SL}(\lambda)+\frac{1}{(4\pi)^2}\left\{-1-\frac{w}{m}\log
\frac{M^2}{\lambda^2}+\log\frac{M^2}{\lambda^2}-\frac{m^2-M^2+s}{2s}\log 
\frac{M^2}{m^2} \right .\nonumber \\
&-& \left . \frac{|\vec{q}\,|}{W}
\log \left( \frac{m^2+M^2-s+2W |\vec{q}\,|}{m^2+M^2-s-2W |\vec{q}\,|}\right) 
\right\}.
\ea
In the former expression we have subtracted the piece 
\be
\frac{w}{(4\pi\,m)^2}\log \frac{M^2}{\lambda^2}
\ee
since this, together with the singular terms stemming from dimensional
regularization:
\be
\hbox{L}(\lambda)=\frac{\lambda^{d-4}}{16\,\pi^2}\left\{\frac{1}{d-4}-
\frac{1}{2}[\ln(4\pi)+\Gamma'+1]\right\}~,
\ee
with $d$ the number of space--time dimensions, 
is reabsorbed in the $\Opt$ HBCHPT counterterms~\cite{nadia}. On the other 
hand, the subtraction constant $\bar{a}^{SL}$ is fixed such that for a value of
$W$ below threshold and in its neighborhood, $g(s)=0$. Note again
that since we require $\bar{a}^{SL}$ to be real, this matching has to
be performed at or below the physical threshold, $s_{\rm thr} = (m+M)^2$.
It is known that the HBCHPT converges best in certain unphysical
regions (but not in all), see e.g. ref.\cite{Paul}.
Therefore, in this energy domain
we have $T_{l\pm}^I= T_{\rm tree+loop}^{I\,l\pm}=T^{I\,l\pm}_{{\rm HBCHPT}}+\Opc$, 
which should be an accurate estimate of the true value of the 
partial wave since around threshold HBCHPT  to $\Opt$ is
convergent. We have
fixed the regularization scale at $\lambda=m$. Note that a change in the scale
$\lambda$ is reabsorbed in a change of the constant $\bar{a}^{SL}$, such that 
at the end the value of $g(s)$ is unaltered.

\section{Results}
\def\theequation{\arabic{section}.\arabic{equation}}
\setcounter{equation}{0}

In this section, we  show and discuss the results. First, we explain
how we have determined the parameters and which of these are strongly
constrained by our fits. Then, we show some representative fits to the
low--lying partial waves, exhibiting in particular the dependence on
some of the coupling constants and mass parameters.

\subsection{Fitting procedure and parameters}
First, since we do not consider inelasticities, we restrict the fits
to a c.m. energy of $W \leq 1.3\,$GeV. We fit the two S-- and 
the four P--waves,
these are $S_{11}, S_{31}$ and $P_{11}, P_{13}, P_{31}$ and $P_{33}$,
respectively. The fits are based on the Karlsruhe--Helsinki KA84 \cite{ka84} and
 the VPI SM95 \cite{sm95} phase shift analyses. These PWAs do not give errors. One
could assume an overall uncertainty of a few percent, but in that way
the smaller partial waves would be too strongly weighted. Also, there
should be some overall systematic error. To deal with these two
effects and to give proper weight to the large partial waves (in
particular the $P_{33}$), we work with an error function for each
partial wave of the form
\be\label{err}
{\rm err(\delta)} = \sqrt{e_s^2 + 0.0025\, \delta^2}~,
\ee
with $e_s=0$ or $e_s=\sqrt{2}$ and the phase shift $\delta$ is given in
radians. In this was a systematic error of $e_s$ and a statistical one of 
a $5\%$ are added in quadrature. As will be shown in more detail below, the 
choice $e_s =0$
gives more weight to the threshold region whereas for $e_s = \sqrt{2}$ we
obtain a very precise description at higher energies, in particular
for the $\Delta$, and an overall better fit for the experimental data up to
$W=1.3$ GeV. Explicit figures for the fits are displayed in the next section.
We now enumerate and discuss the parameters, which are essentially
coupling constants and masses of the excited baryon and meson
resonances. Fitting these parameters can be compared to a
study in CHPT, where one fits the pertinent LECs. Clearly, for very
low energies, this is analogous to assume resonance saturation of the
LECs, which appears to be a sensible approach. However, due to the
explicit propagators,  a fitting procedure as performed here, which is 
also unitary, can be extended to energies
where HBCHPT can not work any more. We have performed various types
of fits, leaving these parameters completely unlimited or 
prescribing limits on some of the parameters based on other phenomenological observations.
These we will discuss in detail in what follows. Clearly, not limiting
certain parameters can lead to rather large (unphysical) values for
them. However, what we mostly can determine are products of couplings
(with respect to some reference mass) and these normalized products
always come out in very limited and sensible ranges. We should also
stress that since our approach does not involve form factors,
one has to be careful to compare our results to the ones obtained
e.g. in meson--exchange models. Having made these general remarks,
consider now the various fits.

\medskip\noindent
First, we discuss the generic results for the case $e_s = \sqrt{2}$.
For the $\Delta (1232)$, we have the
coupling constant and the parameter $Z$. We find that independently of
how we perform the fits, $Z \simeq -0.1$ and $g_{\pi \Delta N} =
2.05$. Note that the values for $Z$ are well within the ranges of
previous determinations, cf. refs.\cite{benn,bkmpi0}.
We note that the standard spin--isospin SU(4)  using the
Goldberger-Treiman relation, $g_{\pi NN} = g_A m/F_\pi$,
leads to (we normalize the coupling in terms of the charged pion mass)
\be
g_{\pi \Delta N} = \frac{3\,g_A}{2\sqrt{2}\, F_\pi} = 2.01/M~.
\ee
This means that the Delta parameters are well determined by the fit.
We have also determined the complex pole in the energy plane
corresponding to the $P_{33}$ resonance and find
\be
(m_\Delta , \Gamma_\Delta /2) = (1210, 53)~\hbox{MeV}~,
\ee
in perfect agreement with dispersive analyses of pion--nucleon
scattering~\cite{Hohler} or pion photoproduction~\cite{Drechsel}.
For the energy range we consider, the Roper only plays a role in the
$P_{11}$ partial wave. We have performed fits with and without Roper.
Including the Roper, we find that for the coupling $\sqrt{R}$,
the fits prefer a
somewhat larger value than the one used in ref.\cite{cucu}.  
Next, we consider the scalar meson exchange. The pertinent
parameters are $\bar{c}_{d,m}, M_S$ and $g_S$. For the constrained
fits, we used $\bar{c}_m = \bar{c}_d = F_\pi /\sqrt{6} \pm 30\%$
and $M_S = 1200\,$MeV~$\pm 30\%$, with the central values discussed 
above. Although these parameters are not
very well constrained by the fits, we observe a few remarkable
features. First, the scalar mass prefers to come out large, which
essentially means that the scalar--isoscalar exchange can be very
well represented by a contact term. Also, the product of the
scalar couplings to the the nucleon and the pions in units of the 
scalar mass squared turns out to be essentially constant,
\be
\frac{ \bar{c}_m \, g_S}{M^2_S} \simeq (1.7 \ldots 2.1) \cdot
10^{-3}~{\rm MeV}^{-1}~.
\ee
This number is about a  factor of two larger than in
ref.\cite{aspects} pointing towards a larger LEC $c_1$ and
consequently a larger pion--nucleon $\sigma$--term. However, it is
known that there are sizeable fourth order corrections to the
scalar sector of (HB)CHPT and we therefore do not want to dwell on
this point any further.
At first sight, these findings are in disagreement with
meson--exchange models or K--matrix approaches which like to
have a light scalar--isoscalar meson. Note, however, that 
due to our chirally symmetric construction, the $S\pi\pi$ coupling
has a different momentum--dependence than what is usually assumed
in these other approaches. To mimic this momentum--dependence,
a light scalar is needed although the resulting fits are much worse in 
such case. For the $\rho$--meson,
we have three parameters. The conventional ones are $g_\rho$ and
$\kappa_\rho$. From these, only $\kappa_\rho$ is reasonably well
determined, $\kappa_\rho = 6.1\pm 0.4$~\cite{MMD}. There is a larger
spread in the values obtained for $g_\rho$ either from the width,
the KFSR relation or fits to the nucleon electromagnetic form factors.
To cope with this uncertainty and remembering again that we have no
form factors at the vertices, in the constrained fits we limit
$g_\rho$ between 5 and 8. For example, the standard
H\"ohler--Pietarinen analysis~\cite{HP} gives $g_{\rho NN} = 5.7$.
Note, however, that even in the fits without
limits, $g_\rho$ always stays in this range. On the contrary, the
value of $\kappa_\rho$ is not well determined by the unconstrained
fits. The additional parameter $T_+$ is always negative and small.
We introduce the dimensionless parameter
$\tilde{T}_+ = (M^2\,M_\rho^2)/(\sqrt{2}\, g_{\rho NN}) \, T_+$,
and find $-0.3 \leq \tilde{T}_+ \leq -0.1$, i.e. the overall contribution of this
additional term is small in most partial waves. In $S_{11}$ and
$P_{13}$, however, varying $\tilde{T}_+$ in the range mentioned leads
to visible changes in the phase shifts, in particular above $W
\simeq 1.2\,$GeV. Therefore, it would be interesting to study the
effect of this extra coupling in other approaches to $\pi$N scattering.
The last parameter concerns the matching of the chiral amplitude to
the unitarized amplitude at the subtraction point $s_0$. Setting
$\sqrt{s_0} = m + aM$, we always find that $a = 0.16$. 

\medskip\noindent
We now turn to the fits with $e_s = 0$. As already remarked, the
$\Delta$ parameters are somewhat less well described, we get
$g_{\pi \Delta N} = 2.25$, $Z$ between $-0.2$ and $-0.1$ and $(m_\Delta , 
\Gamma_\Delta /2) = (1199, 58)$~MeV for the complex pole. On the contrary, the 
coupling $\sqrt{R}$ of the Roper is consistent with the one given in ref. 
\cite{cucu}. The results for the scalar and the $\rho$ meson parameters are 
similar to the ones for $e_s =\sqrt{2}$, again we note that the combination of
the scalar couplings lies in the narrow range of $(1.5 \ldots
1.7)\cdot 10^{-3}\,$MeV$^{-1}$. The matching  of the chiral 
and unitarized amplitudes happens at larger values of $a \simeq 0.5$.
This is, of course, expected and leads to an improved description of
the S--wave scattering lengths as compared to the case of $e_s =\sqrt{2}$.
In the following section, we will tabulate some of the numbers discussed here
and also display some representative fits.

\subsection{Representative fits}

\begin{figure}[htb]
\vspace{-0.7cm}
\centerline{
\epsfysize=5.0in
\epsffile{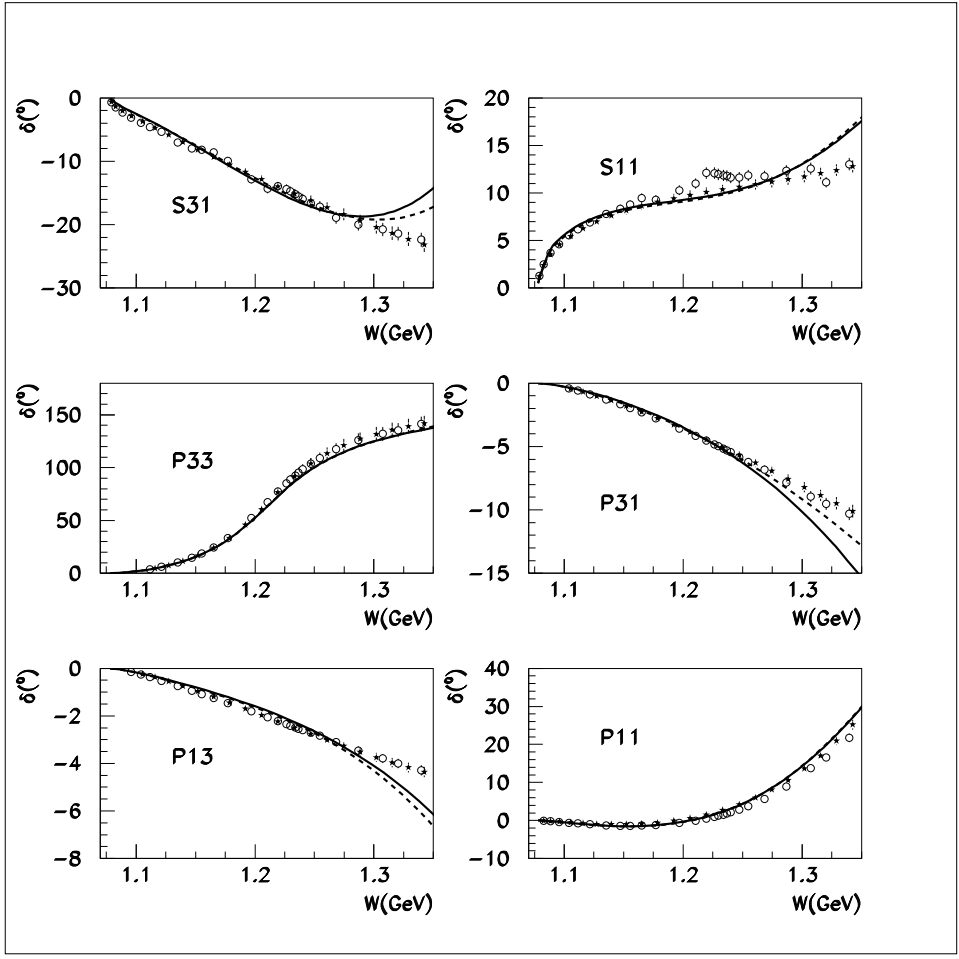}
}
\vskip 0.03cm
\caption{
Fits to phase shifts as a function of the cm. energy $W$
for $e_s = \sqrt{2}$ and $G_V = 53\,$MeV. 
The solid (dashed) lines refer to the constrained (unconstrained)
fits as described in the text.\label{fig:res1}
}

\end{figure}

After having discussed the genuine results in the previous paragraph,
we give in table~\ref{tab:para1} the values of the various fitted
quantities for four  representative fits with $e_s = \sqrt{2}$,
based on the SM95 partial waves (using the KA84
analysis leads to very similar results). For all these fits, the
$\chi^2$/dof is very small. Although for the case of no limits on the 
parameters the $\chi^2$/dof improves, these fits should only be
considered illustrative since they lead to too large parameters in the
scalar sector (we stress again that in such an approach as presented
here one is only sensitive to products of couplings).   
The first fits are based on
limiting the ranges allowed for the scalar couplings $\bar{c}_{d,m}$, 
the scalar mass and the
$\rho$ couplings. In fits~3,4 all these
limits are lifted. Fits~2 and 4 differ from 1 and 3, respectively, in
that the vector coupling $G_V$ is reduced from 67 to 53~MeV (as
explained before). The results are collected in table~\ref{tab:para1}.
The $\chi^2$/dof is of similar
quality in all these fits. Note the remarkable fact that in all the fits, 
also for the case of the second table, the $a$ parameter is always less than 1, 
as required by our final definition of the $g(s)$ function, eq. (\ref{gs2}). 
Note also that in the constrained fits,
the product $\bar{c}_m \, \bar{c}_d$ comes out close to the large
$N_c$ result of (38~MeV)$^2$, cf. eq.(\ref{coupling}), although
individually $\bar{c}_m$ ($\bar{c}_d$) prefers somewhat larger
(lower) values than $F_\pi/\sqrt{6}$. The scalar mass always
tends to the upper end of the given range. For these four fits, we find
$\bar{c}_m \, g_S/ M^2_S = (2.1, 1.9, 1.7, 1.7) \cdot
10^{-3}\,$MeV$^{-1}$, in order.
If we, however, fix the scalar mass to 1200~MeV, $\bar{c}_d$ comes out
closer to the large $N_c$ prediction and $g_S$ is smaller, but again
we find that the aforementioned product of scalar couplings is
$1.7\cdot 10^{-3}\,$MeV$^{-1}$. As noted before, our fits are only
sensitive to the products $G_V g_\rho$, $G_V \kappa_\rho$ and $G_V
T_+$. This is reflected by the fact that for fits 3 and 4, we find
the same value for $G_V g_\rho$ and almost the same for fits 1 and 2.
The corresponding S-- and P--wave phase shifts are shown in fig.\ref{fig:res1}.

\renewcommand{\arraystretch}{1.2}
\begin{table}[H]
\begin{center}
\begin{tabular}{||l||c|c|c|c||}
\hline
                   & Fit~1   &  Fit~2   &   Fit~3   &  Fit~4   \\
\hline\hline
$g_{\pi\Delta N}$  &  2.05     &   2.04      &    2.05     &   2.05  
                   \\
\hline
$Z$                &  $-$0.16 &  $-$0.08    &   $-$0.05   &  $-$0.05
                   \\
\hline
$\sqrt{R}$         &  0.67     &   0.79  &    0.79     &  0.79
                   \\
\hline
$\tilde{T}_+$      &  $-$0.15  &  $-$0.26    &   $-$0.09   & $-$0.09 
                   \\
\hline     
$\bar{c}_{d}$~[MeV]&  26.2~(25.,50.) & 25.0~(25.,50.) &  5.9$\cdot 10^5$ & 115.8
                   \\
\hline
$\bar{c}_{m}$~[MeV]&  50.0~(25.,50.) & 43.9~(25.,50.) & 1.0$\cdot 10^6$ & 199.8
                   \\
\hline
$M_S$~[MeV]        & 1560~(840,1560) & 1560~(840,1560)& 3.8$\cdot 10^7$ & 63681.
                   \\  
\hline
$g_S$              &  100.2$^\dagger$ & 104.3$^\dagger$ & 2.4$\cdot 10^6$$^\dagger$ &
  35159.$^\dagger$
                   \\ 
\hline
$G_V$~[MeV]        &  67$^*$   &   53$^*$    & 67 $^*$ &  53$^*$
                   \\
\hline
$g_\rho$           &  5.00~(5,8) & 5.00~(5,8) & 6.08&  7.70  \\
\hline
$\kappa_\rho$      &  5.70~(5.7,6.5) & 5.70~(5.7,6.5) & 3.53  &  3.53 \\
\hline
$a$                &  0.16  & 0.16  & 0.16 & 0.16  \\
\hline
$\chi^2$/dof       &  21.5/355 &  15.6/355 & 12.9/355 & 12.0/355 \\
\hline\hline
\end{tabular}
\caption{Representative fits with $e_s = \sqrt{2}$ as explained in the text. The $^*$
         denotes a fixed quantity and the $^\dagger$ indicates that $g_S$ appears 
         always multiplying $\bar{c}_{d,m}$. For fits~1,2, the ranges on the
         various parameters as explained in the text are given in the
         round brackets.
\label{tab:para1}} 
\end{center}
\end{table}

\renewcommand{\arraystretch}{1.2}
\begin{table}[H]
\begin{center}
\begin{tabular}{||l||c|c|c|c||}
\hline
                   & Fit~1   &  Fit~2   &   Fit~3   &  Fit~4   \\
\hline\hline
$g_{\pi\Delta N}$  &  2.21     &   2.25      &    2.26     &   2.26  
                   \\
\hline
$Z$                &  $-$0.18 &  $-$0.12    &   $-$0.09   &  $-$0.09
                   \\
\hline
$\sqrt{R}$         &  0.46     &   0.56  &    0.54     &  0.54
                   \\
\hline
$\tilde{T}_+$      &  $-$0.19  &  $-$0.23    &   $-$0.08   & $-$0.08 
                   \\
\hline     
$\bar{c}_{d}$~[MeV]&  25.~(25.,50.) & 25.7~(25.,50.) &  5.7$\cdot 10^5$ & 87847.
                   \\
\hline
$\bar{c}_{m}$~[MeV]&  45.2~(25.,50.) & 45.3~(25.,50.) & 1.0$\cdot 10^6$ & 126.7
                   \\
\hline
$M_S$~[MeV]        & 1560~(840,1560) & 1560~(840,1560)& 3.9$\cdot 10^7$ & 222.5
                   \\  
\hline
$g_S$              &  92.6$^\dagger$    & 87.9$^\dagger$ & 
                   2.3 $\cdot 10^6$$^\dagger$ &  $53682.^\dagger$
                   \\ 
\hline
$G_V$~[MeV]        &  67$^*$   &   53$^*$    & 67 $^*$ &  53$^*$
                   \\
\hline
$g_\rho$           &  5.00~(5,8) & 5.62~(5,8) & 6.76  &  8.55  \\
\hline
$\kappa_\rho$      &  5.7~(5.7,6.5) & 5.7~(5.7,6.5) & 3.64  &  3.64 \\
\hline
$a$                &  0.65  & 0.41  & 0.51 & 0.51  \\
\hline
$\chi^2$/dof       &  1.15 &  1.00 & 0.78 & 0.78 \\
\hline\hline
\end{tabular}
\caption{Representative fits with $e_s = 0$ as explained in the text. The $^*$
         denotes a fixed quantity and the $^\dagger$ indicates that $g_S$ appears 
         always multiplying $\bar{c}_{d,m}$. For fits~1,2, the ranges on the
         various parameters as explained in the text are given in the
         round brackets.
\label{tab:para2}} 
\end{center}
\end{table}

\noindent
We point out that in particular the $P_{11}$ and the $P_{33}$ partial 
waves are well described, and a good overall description of data up to $W=1.3$ GeV is 
achieved. If one calculates the S--wave scattering lengths and the
P--wave scattering volumes, these come out in reasonable agreement
with the data. 

%

\begin{figure}[H]
\centerline{
\epsfysize=5.0in
\epsffile{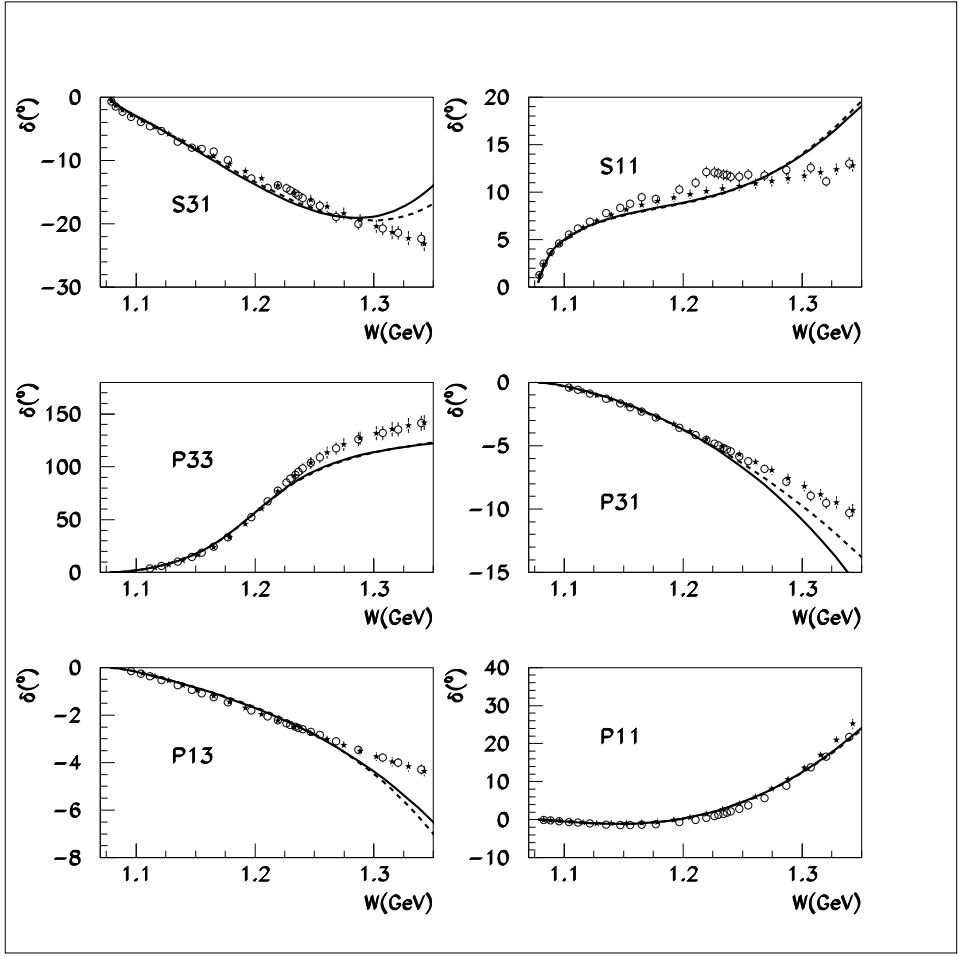}
}
\vskip 0.3cm
\caption{
Fits to phase shifts as a function of the cm. energy $W$
for $e_s = 0$ and $G_V = 53\,$MeV. 
The solid (dashed) lines refer to the constrained (unconstrained)
fits as described in the text.\label{fig:res2}
}

\end{figure}

\medskip\noindent
We now turn to the fits with $e_s = 0$. We have performed
the same four fits as described above. The resulting 
parameters are collected in table~\ref{tab:para2}. These are
similar to what was found before. The most striking differences
are the Roper coupling, which comes out close to the value derived
in ref.\cite{bora} and the larger value for the matching parameter
$a$. Still, we find $a < 1$, as demanded by our definition of the
matching function $g(s)$. The resulting $\chi^2$/dof is around 1. Note that this
number is only of limited meaning since it strongly depends on the choice of $e_s$, 
which now is 0 and for table \ref{tab:para1} it was $\sqrt{2}$, and on the 
assumed statistical uncertainty.
The resulting S-- and P--wave scattering lengths (volumes)
are collected in table~\ref{tab:thr}. With the exception of
$a_{0+}^+$, they are in good agreement with the results based
upon the Karlsruhe and VPI partial wave analysis. We stress 
that $a_{0+}^+$ is very sensitive to cancelations between 
contact term and loop contributions, which are individually much
bigger than the total sum~\cite{bkmpin}, and thus is only expected to be described
precisely at fourth order. The resulting phase shifts for $W \leq
1.3\,$GeV are shown in fig.\ref{fig:res2}. They are slightly worse
than the ones for the fits with $e_s = \sqrt{2}$. In particular, the
$P_{33}$ partial wave is not as precisely described as before, but
still the complex pole is only slightly displaced as compared to
the fits with $e_s = \sqrt{2}$. 

\begin{table}[hbt]
\begin{center}
\begin{tabular}{|c||c|c|c||c|c|}
    \hline
   Obs.    &  Fit 1  &  Fit 2  & Fit 3 & KA85 & SP98 \\
    \hline\hline
$a^+_{0+}$  & $1.59$ & $1.13$ & $0.87$    
      & $-0.83$ & $ 0.0 \pm 0.1$ \\
$a^-_{0+}$  & $8.37$ & $8.31$ & $8.28$   
      & $ 9.17$ & $ 8.83 \pm 0.07$ \\
$a^+_{1-}$  & $-4.92$ & $-4.85$ & $-5.03$    
      & $-5.53$ & $-5.33 \pm 0.17$ \\ 
$a^+_{1+}$  & $ 13.71$ & $13.87$ & $ 13.73$  
      & $ 13.27$  & $ 13.6 \pm 0.1$ \\
$a^-_{1-}$  & $-1.23$ &  $-1.27$ & $-1.21$  
      & $-1.13$ & $-1.00 \pm 0.10$ \\
$a^-_{1+}$  & $-8.08$ & $-8.15$ & $-8.10$  
      & $-8.13$ &  $-7.47 \pm 0.13$ \\
  \hline\hline
  \end{tabular}
  \caption{Values of the S-- and P--wave threshold parameters for the various 
  fits as described in the text in comparison to the respective
           data (taken from ref.\protect{\cite{nadia}}).
           Fits~3 and 4 give identical results within the precision displayed. 
           Units are appropriate inverse powers of the pion mass times 10$^
           {-2}$.
           \label{tab:thr}}
\end{center}
\end{table}

\newpage

\section{Conclusions}
\def\theequation{\arabic{section}.\arabic{equation}}
\setcounter{equation}{0}

In this paper, we have developed a novel chiral unitary approach 
for meson-baryon processes. As a first application, we have
considered  elastic pion--nucleon scattering from threshold up to
the opening of the inelastic channels at $W \simeq 1.3\,$GeV.
Making use of the time--honored $N/D$ method,
we have derived the most general structure of any pion--nucleon
partial wave amplitude neglecting unphysical cuts. The latter are
then treated in a perturbative chiral loop expansion 
based on the power counting of heavy
baryon chiral perturbation theory. The most important and novel
features and results of this approach can be summarized as follows:
\begin{enumerate}
\item[i)] The approach is based on subtracted dispersion relations and
  is by construction relativistic. No form factors, finite momentum
  cut--offs  or regulator
  functions are needed to render the resummation finite. This is
  done by a subtraction of the dispersion relations below the physical threshold.
\item[ii)] The amplitude is matched to the one of (heavy baryon)
  chiral perturbation theory at third order. This is done at an energy
  slightly below the physical threshold. In this region, the HBCHPT
  amplitude is expected  to converge. It is straightforward to extend
  this matching to a fourth order and/or a relativistic CHPT amplitude.
\item[iii)] The explicit resonance degrees of freedom contain free
   parameters (couplings and masses) which are determined from a best
   fit to the pion--nucleon partial waves. Some of these parameters,
   in particular the ones of the $\Delta$, are very well determined.
   Couplings of the scalar and vector mesons can be pinned down less
   precisely. We have also pointed out a new coupling of the $\rho$
   not considered in conventional schemes. Our approach is consistent with 
   the concept of resonance saturation of the low--energy constants.
\item[iv)] The pion--nucleon phase shifts are well described 
   below the onset of inelasticities. With the exception of the 
   isoscalar S--wave scattering lengths, we also find a reasonable
   description of the $\pi$N threshold parameters. This can be
   improved by matching to a more precise (HB)CHPT amplitude.
\item[v)]Our method is very general. It embodies (as a particular
  limiting case) the inverse amplitude method (IAM) \cite{IAM} recently 
  applied to the $\pi$N scattering in \cite{ramonet}. For a discussion about
  this point see refs. \cite{WW,report}. The chiral unitary 
  approach \cite{valencia}, based on the lowest order 
  tree level meson--baryon interactions, appears also as a
  special case \cite{report}.
\item[vi)] The extension of this approach to coupled channels (like
  $\pi N \to \eta N$) and to the three--flavor sector (with
  coupled channels) is in principle straightforward. For the
  kaon--nucleon system, the matching to the chiral perturbation theory
  amplitude is more tricky due to the appearance of subthreshold resonances. 
  Work along these lines is in progress.
\end{enumerate}

\bigskip

\subsection*{Acknowledgements}
We are grateful to Nadia Fettes for some clarifying comments.
This work was supported in part by the EEC-TMR program under
contract no. ERBFMRX-CT98-0169.

\newpage

\end{document}